\documentstyle[12pt,epsf,doublespace]{article}

\def\beq{\begin{equation}}
\def\eeq{\end{equation}}
\def\beqa{\begin{eqnarray}}
\def\eeqa{\end{eqnarray}}

\newcommand{\bfp}{\mbox{\boldmath $p$}}
\newcommand{\bfq}{\mbox{\boldmath $q$}}
\newcommand{
\xtdots}{\hspace{3pt}\ddot{}\mbox{}\hspace{3.3pt}\dot{}\hspace{-6pt}x}

\begin{document}
\thispagestyle{empty}

\mbox{}
\vspace{1cm}
\begin{center}
{\bf THE USE OF HAMILTONIAN MECHANICS IN} \\
\vspace{0.5cm}
{\bf SYSTEMS DRIVEN BY COLORED NOISE} \\
\vspace{0.5cm}
{\it S.J.B.Einchcomb and A.J.McKane} \\
\vspace{0.5cm}
Department of Theoretical Physics \\
University of Manchester \\ Manchester M13 9PL, UK \\
\end{center}

\begin{abstract}
The evaluation of the path-integral representation for stochastic processes
in the weak-noise limit shows that these systems are governed by a set of
equations which are those of a classical dynamics. We show that, even when
the noise is colored, these may be put into a Hamiltonian form which leads
to better insights and improved numerical treatments. We concentrate on
solving Hamilton's equations over an infinite time interval,
in order to determine the leading order contribution to the mean escape time
for a bistable potential. The paths may be oscillatory and inherently
unstable, in which case one must use
a multiple shooting numerical technique over a truncated time
period in order to calculate the infinite time optimal paths to a given
accuracy.  We look at two systems in some detail: the underdamped Langevin
equation
driven by external exponentially correlated noise and the overdamped Langevin
equation driven by external quasi-monochromatic noise. We deduce that
caustics, focusing and bifurcation of the optimal path are general
features of all but the simplest stochastic processes.
\end{abstract}

\newpage
\pagenumbering{arabic}

\section{Introduction}

The subject of noise induced activation has received a great deal of attention
in the last decade or so with the development of new techniques
which allow systems where the noise is not white (i.e.colored) to be
studied in a systematic and controlled way \cite{bm},\cite{bmn}. While these
techniques were being refined it was natural that only the simplest of systems
were studied: those which consisted of a single particle moving in a
one-dimensional potential with a relatively simple form of noise. More
recently the investigation of models acted upon by white noise and having more
than one degree of freedom has revealed novel effects such as caustics and
focusing singularities \cite{ms1}-\cite{cds}. In these systems it was also
found that the leading order term in the expression for the mean escape time,
that is the action, was reduced unexpectedly \cite{ms1}.

The appearance of such features can be understood in the following way. In the
limit of weak noise the dynamics of the system is governed by a set of
equations which are the extrema of the action in the path-integral formulation
of the stochastic process. These equations have the same form as those of
Newtonian mechanics, if the original stochastic dynamics was underdamped and
the noise was white. Hence the dynamics of these stochastic processes is
controlled by trajectories in a $2n$-dimensional phase space, where $n$ is the
number of degrees of freedom of the system. Placed in this context, it is not
unnatural to expect caustics and
focusing singularities \cite{arn}. In this paper we will show that similar
effects will also occur in systems with one degree of freedom, but with a
more complicated type of noise. We do this by using a generalized
Hamiltonian formalism to show that a phase space can be constructed which is
multi-dimensional and hence, since these systems do not satisfy detailed
balance, will be expected to show the phenomena mentioned above.

Our starting point is the observation that a system consisting of a single
degree of freedom, but acted upon by a rather general form of external
noise \cite{ajm}, can be written as a Markov process
which consists of a number of equations, only one of which involves a noise
term (which is white). These equations can be combined into a single equation,
at the expense of introducing higher time derivatives. Essentially we have
traded a simple system acted upon by a complicated noise term, for a
complicated system acted upon by a simple white noise. We shall give explicit
examples later in this paper. A process of the kind we have been describing
can be defined by the generic stochastic differential equation
\beq\label{eq:gen}
f(x,\dot{x},\ddot{x},\ldots,x^{(n)};t)=\eta(t)
\eeq
where $\eta(t)$ is Gaussian white noise of strength $D$.

Let us now make the above comments on the emergence of a classical dynamics
in the weak noise limit more concrete by outlining how a
path-integral representation for the conditional probability distribution of
a process defined by (\ref{eq:gen}) can be written down \cite{mbl}. One begins
by using (\ref{eq:gen}) to transform the probability density functional for
white noise given by
\beq\label{eq:pdfw}
P[\eta]={\cal C}\exp\left(-\frac{1}{4D}\int_{t_0}^{t}\eta^{2}(t)dt
\right)
\eeq
to the probability density functional for that of the coordinate $x$:
\beq\label{eq:pdfx}
P[x]={\cal N}J[x]\exp\left( -\frac{S[x]}{D} \right)
\eeq
where $S[x]$ is the action mentioned above and $J[x]$ is the Jacobian of the
transformation. The action is so called since it may be written as
\beq \label{eq:4.2}
S[x]=
\int_{t_0}^{t}dt~L(x,\dot{x},\ddot{x},\ldots,x^{(n)};t)
\eeq
where $L(x,\ldots)$ given by
\beq\label{eq:lag}
L(x,\dot{x},\ddot{x},\ldots,x^{(n)};t)=\frac{1}{4}
[f(x,\dot{x},\ddot{x},\ldots,x^{(n)};t)]^{2}
\eeq
This has the form of a Lagrangian for a mechanical system, if the noise is
white and the motion is overdamped, so that no time-derivatives higher than the
first appear in (\ref{eq:gen}). For colored noise processes of the type we are
investigating here, there are higher time-derivatives in the Lagrangian and
the analogy is now with a generalized form of mechanics. The precise form of
the Jacobian factor will not be required in this paper since we will be
performing our calculations to leading order only, and the Jacobian only enters
at next order. Probability distributions, correlation functions and other
quantities of interest can be found by integration of the appropriate
functions over paths $x(t)$ with weight (\ref{eq:pdfx}).  In the limit of
$D\rightarrow 0$ these path-integrals can be evaluated by the method of
steepest descents; the paths which dominate the integrals being the ones for
which $\delta S[x]/\delta x=0$.
This leads to the Euler-Lagrange equation for the optimum path which will be
in general a $2n^{th}$ order non-linear differential equation given by
\beq\label{eq:ele}
\sum_{i=0}^{n} (-1)^{n} \frac{d^{n}}{dt^{n}} \left( \frac{\partial L}
{\partial x^{(n)}} \right)=0
\eeq
In general, this equation will have no analytical solution and one has to
rely on numerical techniques \cite{bmn,nbm}. A numerical solution will involve
the decomposition of such an equation into $2n$ coupled first order non-linear
differential equations.  It would be convenient to derive the an expression for
the optimal path in such a format automatically.  This is instantly provided
by using Hamilton's formalism as an alternative to the Lagrangian method.
Another advantage of using this method is that the greater geometrical
structure of Hamiltonian mechanics gives one a better insight into why the
optimum paths take on the particular form that they do.

The outline of the paper is as follows. In section two we construct the
generalized Hamiltonian formalism appropriate to problems of this type
and give the case of the underdamped Langevin equation driven by white
noise as an example. In section 3 we use the formalism to find the
mean first passage time for this underdamped problem, but now with
exponentially correlated noise; a task which could not be achieved using
the Lagrangian formalism \cite{nbm}. The case of quasi-monochromatic noise
is discussed in section 4 and we conclude in section 5.

\section{Hamiltonian Formalism}

For a dynamical system which is defined by a Lagrangian of the form
$L(x,\dot{x},\ddot{x},\ldots,x^{(n)};t)$, a Hamiltonian structure can still
be constructed (see, for instance, \cite{whi}). To do so one introduces a
generalized coordinate vector $\bfq$ spanning an $n$
dimensional space with components $\{q_{1},\ldots,q_{n}\}$ such that
\beq\label{eq:4.4}
q_{i}=x^{(i-1)}
\eeq
and one writes
\beq\label{eq:4.4a}
L(x,\dot{x},\ddot{x},\ldots,x^{(n)};t)=\sum_{i=1}^{n}p_{i}\dot{q}_{i}
- H({\bfq},{\bfp};t)
\eeq
where the $p_i$'s have yet to be defined. Now if one demands that
$\dot{p}_i = -\partial H/\partial q_i$, it follows from (\ref{eq:4.4a})
that
\beq\label{eq:4.47}
\frac{\partial L}{\partial x^{(j)}} = p_j + \dot{p}_{j+1} \ \ \ ; j=0,...,n
\eeq
where $p_0$ and $p_{n+1}$ are defined to be zero. From (\ref{eq:ele}) and
(\ref{eq:4.47}) one sees that the components $\{p_{1},\ldots,p_{n}\}$
of the generalized momentum vector $\bfp$ should be taken to be
\beq\label{eq:4.5}
p_{i}=\sum_{j=i}^{n} (-1)^{j-i} \frac{d^{j-i}}{dt^{j-i}}\left( \frac{\partial
L}{\partial x^{(j)}} \right)
\eeq
Hence, by construction, the optimum path given by the $2n^{th}$ order
differential equation (\ref{eq:ele}) can also be found by solving the $2n$
first order differential equations
\beq\label{eq:hamq}
\dot{q}_{i}=\frac{\partial H}{\partial p_{i}}
\eeq
and
\beq\label{eq:hamp}
\dot{p}_{i}=-\frac{\partial H}{\partial q_{i}}
\eeq
If the Lagrangian does not involve time explicitly, then the Hamiltonian, $H$,
also has no explicit time dependence, and since $dH/dt = \partial H/\partial
t$,
the Hamiltonian is conserved. This reduces by one the number of integrals that
have to be performed.

As an example we shall consider the underdamped Langevin equation driven by
white noise
\beq\label{eq:4.9}
m\ddot{x}+\alpha\dot{x}+V'(x)=\eta(t)
\eeq
where $\eta(t)$ is Gaussian white noise of strength $D$. Here $V(x)$ is
assumed to be a double well potential and $\alpha$ is a friction constant
which will be set equal to unity by an appropriate choice of units of time.
The Lagrangian for this process is given by
\beq\label{eq:4.10}
L(x,\dot{x},\ddot{x})=\frac{1}{4}(m\ddot{x}+\dot{x}+V'(x))^{2}
\eeq
and the Hamiltonian is found to be
\beq\label{eq:4.11}
H({\bf q},{\bf p})=p_{1}q_{2}+\frac{p_{2}^{2}}{m^{2}}
-\frac{p_{2}}{m}\left( q_{2}+V'(q_{1}) \right)
\eeq
The optimum path is then the solution of Hamilton's equations:
\[
\dot{q}_{1}=q_{2}
\]
\[
\dot{q}_{2}=\frac{2p_{2}}{m^{2}}-\frac{q_{2}}{m}-\frac{V'(q_{1})}{m}
\]
\[
\dot{p}_{1}=\frac{p_{2}V''(q_{1})}{m}
\]
\beq\label{eq:4.12}
\dot{p}_{2}=\frac{p_{2}}{m}-p_{1}
\eeq
and the action is given by
\beq\label{eq:4.12a}
S=\int_{t_0}^{t} \frac{p_{2}^{2}}{m^{2}}~dt
\eeq
In this case we can find the required solutions explicitly enough to allow
us to write down the action in closed form. We are searching for solutions
which begin at extrema of the potential with all time derivatives of the
coordinate equal to zero. This immediately tells us that $H=0$ for these
solutions, which is a common feature in models of this type. For this simple
case there are only two of these solutions: a ``downhill" solution given by
$m\ddot{x}+\dot{x}+V'(x)=0$ and an ``uphill" solution given by
$m\ddot{x}-\dot{x}+V'(x)=0$ \cite{nbm}. These solutions can easily be found
as $H=0$ solutions of (\ref{eq:4.12}): the downhill solution has
$p_{1}=p_{2}=0$ and zero action and the uphill solution has $p_{1}=V'(q_{1})$,
$p_{2}=mq_{2}$ and action given by
\beq\label{eq:smwh}
S=\left[ \frac{1}{2}mq_{2}^{2}+V(q_{1})\right]_{t_{0}}^{t}
\eeq
The interpretation of (\ref{eq:smwh}) depends on exactly what quantity is being
calculated.  For example if one wished to find the stationary probability
distribution then one would take $t_{0}\rightarrow -\infty$ so that
$S=m\dot{x}^{2}/2+V(x)$ in terms of the original variable $x$, which just gives
the Maxwell-Boltzmann distribution. On the other hand, if one wished to find
the mean escape rate from a potential well, one is interested in paths which
take an infinite time to interpolate between stable and unstable points of the
potential and are at rest at both ends. This gives $S=\Delta V$, where $\Delta
V$ is the barrier height.  For the rest of this paper we will restrict
ourselves to the calculation of this quantity and so will assume an
infinite time interval in what follows. Since for colored noise processes,
which are the real interest of this paper, we cannot, in general, calculate the
action explicitly, we will choose the specific double-well potential
\beq\label{eq:v}
V(x)=-\frac{x^{2}}{2}+\frac{x^{4}}{4}
\eeq
to illustrate our techniques. If we choose to investigate activation from the
left-hand well to the right-hand well, then the section from zero to plus one
will be a downhill path with zero action. Thus we need only concern ourselves
with the section of the path from minus one to zero. Having illustrated the
technique on a simple white noise problem we now go on to investigate the
same system, but acted upon by exponentially correlated noise.

\section{Exponentially correlated noise}
In this section we consider the process modelled by the Langevin equation
\beq\label{eq:4.13}
m\ddot{x}+\dot{x}+V'(x)=\xi(t)
\eeq
where $\xi(t)$ is Gaussian colored noise whose correlation function is given by
\beq\label{eq:4.14}
\langle \xi(t)\xi(t') \rangle =\frac{D}{\tau}\exp\left( -\frac{|t-t'|}{\tau}
\right)
\eeq
This represents the simplest generalization of the noise in the system
first investigated by Kramers \cite{kra}, which it reduces to in the
$\tau\rightarrow 0$ limit. One can see that it is the simplest generalization
by replacing (\ref{eq:4.14}) by the condition that $\xi$ obeys the first order
differential equation
\beq\label{eq:4.15}
\tau\dot{\xi}+\xi=\eta(t)
\eeq
where $\eta$ is a Gaussian white noise of strength $D$. Equations
(\ref{eq:4.13})
and (\ref{eq:4.15}) form an equivalent Markov process with two degrees of
freedom. These equations may be combined into the single third-order
stochastic differential equation
\beq\label{eq:4.15a}
m\ddot{x}+\dot{x}+V'(x)
+\tau (m\xtdots+\ddot{x}+\dot{x}V''(x)) = \eta (t)
\eeq
This is of the form (\ref{eq:gen}) with $n=3$ and so we expect to be able
to describe the weak-noise limit of this system using either Lagrangian or
Hamiltonian dynamics.

The Lagrangian approach to this problem has been investigated by Newman
et al \cite{nbm}. However these authors were only able to explore the
dynamics of the system for relatively small masses; they were unable to
analyse the underdamped regime. We shall show in this section that the
Hamiltonian approach allows us to do this.

Using (\ref{eq:lag}) and (\ref{eq:4.15a}) we can write down a Lagrangian
for this system:
\beq\label{eq:4.16}
L(x,\dot{x},\ddot{x},\xtdots) = \frac{1}{4}
\left[ (m\ddot{x}+\dot{x}+V'(x))
+\tau (m\xtdots+\ddot{x}+\dot{x}V''(x)) \right]^{2}
\eeq
The equivalent Hamiltonian is found in the way described in section 2 to be:
\beq\label{eq:4.17}
H(\bfq,\bfp)=p_{1}q_{2}+p_{2}q_{3}
+\frac{p_{3}^{2}}{m^{2}\tau^{2}}
-\frac{p_{3}}{m\tau}(mq_{3}+q_{2}+V'(q_{1}))
-\frac{p_{3}}{m}(q_{3}+q_{2}V''(q_{1}))
\eeq
with Hamilton's equations given by (\ref{eq:hamq}) and (\ref{eq:hamp}).
The action reduces to
\beq  \label{eq:4.19}
S=\int_{-\infty}^{\infty}dt \frac{p_{3}^{2}}{m^{2}\tau^{2}}
\eeq
We now wish to find the value of the infinite time action for the bistable
potential (\ref{eq:v}). The downhill solution is, as usual, trivial:
$p_{n}=0$; $n=1,2,3$, which gives zero action. The uphill solution cannot be
found analytically for general $m$ and $\tau$ and only perturbative methods
and numerical solutions are available. The $m=0$, general $\tau$ problem is
extensively discussed in \cite{bmn}, along with perturbative expansions in
the small $\tau$ and large $\tau$ regimes. Therefore we will restrict ourselves
to $m>0$. In Ref \cite{nbm} a numerical calculation of the action for certain
values of $m$ and $\tau$ have been given, as well as perturbation expansions
for small $m$ and small $\tau$. In the rest of the section we will expand on
this treatment, extending it and investigating the previously unexplored
underdamped regime.

For general $m$, but small $\tau$, the action for the uphill path has the
form
\beq\label{eq:4.20}
S(m,\tau)=S_{0}+\tau^{2}S_{1}(m)+~O(\tau^{4})
\eeq
where $S_{0}=1/4$ is the white noise action for this potential. The first
correction $S_{1}(m)$ has the simple form \cite{nbm}
\beq\label{eq:4.21}
S_{1}(m)=\int_{-\infty}^{\infty} \ddot{x}_{0}^{2}~dt
\eeq
where $x_{0}$ is the optimal path for white noise ($\tau=0$) and is given by
the non-linear differential equation
\beq\label{eq:4.22}
m\ddot{x}_{0}-\dot{x}_{0}+V'(x_{0})=0
\eeq
with the boundary conditions $x_{0}(-\infty)=-1$ and $x_{0}(\infty)=0$.

In the paper by Newman et al \cite{nbm} this quantity was calculated for small
$m$ only. However equation (\ref{eq:4.22}) is stable if integrated backwards
in time, i.e. starting at $x_{0}=0$ going to $x_{0}=-1$ (stability is discussed
further when the full solution is considered). Hence, one can use a simple
initial value integrating scheme, such as a fourth-order Runga Kutta, starting
with an infinitely small velocity $\dot{x}_{0}=\delta$. While in a formal
sense the path over the infinite time interval is only found in the limit
$\delta\rightarrow 0$, in practice we find that if $\delta $ is small, the
value of the action does not depend on it. The results
for $S_{1}(m)/S_{0}$ are plotted in figure 1 as a function of $\log_{10}(m)$.
The dotted line shows a seventh-order perturbative calculation of $S_{1}(m)$:
\beq\label{eq:4.23}
S_{1}(m)=S_{0}\left(\frac{1}{2}+\frac{m}{5}-\frac{m^{2}}{5}-\frac{2m^{3}}{5}
+\frac{3m^{4}}{10}+\frac{9m^{5}}{5}-\frac{3m^{6}}{5}-\frac{778m^{7}}{55}
\right)+~O(m^{8})
\eeq
Figure 1 shows the excellent agreement between the series (\ref{eq:4.23}) and
the numerical solution for $m$ less than about $0.3$, and the catastrophic
failure of the series above that value. This breakdown of perturbation theory
may be due to the change in the nature of the solutions that occurs at $m=1/8$
(see below) and a calculation of more terms in the series
(\ref{eq:4.23}) might show that the value of $m$ at which the breakdown
occurs approaches the value $0.125$. This figure also shows that the
value of $S_{1}(m)$ has a maximum when plotted against $\log_{10}(m)$. Such
maxima are also seen when plotting actions against $\log_{10}(m)$ (\cite{nbm}
and Figure 2 below). The existence of these maxima are a consequence of the
non-linear nature of the problem.

Now let us go on to a numerical study of the solution of Hamilton's equations
for general $m$ and $\tau$. As a first step we linearize the equations about
the endpoints $q_{1}=a$, where $a$ is zero or minus one. To do this we
approximate the potential by a parabola
$V(q_{1}) = V(a)+\frac{1}{2}V''(a)(q_{1}-a)^{2}$, which leads to linear
Hamilton's equations with solutions of the form
$q_{1}=a+\sum_{n} A_{n}e^{\lambda_{n}t}$ where $A_{n}$ are arbitrary constants
specified by the boundary conditions. The $\lambda_{n}$ have six possible
values:
\beq\label{eq:4.26}
\lambda_{n}=\pm\frac{1}{\tau}~,~\pm\left(\frac{1}{2m}\pm
\frac{\sqrt{1-4mV''(a)}}{2m} \right)
\eeq
When $a=0$, $V''(a)=-1$ and we require that $q_{1} \rightarrow 0$ as
$t \rightarrow \infty$ for an uphill path, so we select only those
$\lambda _n$ which are negative. Conversely, when $a=-1$, $V''(a)=2$ and  we
require that $q_{1}\rightarrow -1$ as $t \rightarrow -\infty$, thus we must
only take values of $\lambda_{n}$ which are positive, i.e.
\beq\label{eq:4.28}
\lambda_{1,2,3}=\frac{1}{\tau}~,~
\left(\frac{1}{2m}+\frac{\sqrt{1-8m}}{2m} \right)~,~
\left(\frac{1}{2m}-\frac{\sqrt{1-8m}}{2m} \right)
\eeq
If $m<1/8$ no problem arises. However if $m>1/8$, two of the quantities in
(\ref{eq:4.28}) are complex which is a signal that the solutions may be
oscillatory. Actually if $\tau >2m$ the real one dominates, implying that the
solutions are not oscillatory near to the stable fixed point in this case.
In summary, we can say that, as in the underdamped Langevin equation with
white noise, the system oscillates about the bottom of the potential wells
before making a transition, unless $m<1/8$ or $m>1/8$ and $\tau >2m$. In
this case the substitution $y(x)=\dot{x}(t)$, which was the basis of the
approach in \cite{nbm}, fails and another technique has to be used.

The advantage of solving for $y(x)$ is that since $-1\leq x \leq 0$, the
differential equation has to be solved in a finite range. Unfortunately, in
the region of parameter space where the oscillatory solution exists we have
to solve Hamilton's equations over the range $-\infty <t<\infty$. In
practice, of course, we have to truncate this span to a large, but finite,
value $T$ and use the boundary conditions
\[
q_{1}(- \frac{T}{2} )= -1 \hspace{2cm} q_{1} (\frac{T}{2} )=0
\]
\[
q_{2}(\pm \frac{T}{2} )=0
\]
\beq\label{eq:4.34}
q_{3}(\pm \frac{T}{2} )=0
\eeq
and calculate the action
\beq\label{eq:4.35}
S=\int_{-\frac{T}{2}}^{\frac{T}{2}} \frac{p_{3}^{2}}{m^{2}\tau^{2}}~dt
\eeq
The extrapolation $T \rightarrow \infty$ is, in fact, not a problem; the
actual transition happens over a time scale of order a few $m$ and
for the majority of the time the particle is almost at rest at the two
endpoints. Furthermore, this decay of the position, velocity, etc, at the
endpoints is exponential, which means that the truncation can be carried out
extremely accurately.
However the simple initial value techniques of solution which are used in
shooting routines can no longer be used, as this problem is inherently
unstable. This is because, as one can see from (\ref{eq:4.26}), there are
growing solutions at both the end points. In analytic treatments these
solutions can be ignored by setting the arbitrary constants ($A_{n}$) to
zero. Numerically, roundup errors introduced either through machine precision
or through the solution algorithm, make these constants small, but non-zero.
Since these solutions grow exponentially, whereas the required solution decays
exponentially, they soon take over, and any hope of solving the problem
numerically by this method is destroyed.

Normally, it is possible to solve inherently unstable problems by using a
relaxation technique or collocation such as COLSYS \cite{col}. However, as
noticed in Ref \cite{nbm} this technique is poorly convergent when the
solution is oscillatory. Instead, one can attempt to proceed using either
invariant embedding or multiple shooting techniques. It is the latter that we
have used; calculating the action for $\tau<2m$ using MUSN \cite{mus}. These
techniques damp out the exponentially growing solution by splitting the total
time span into several smaller time segments and then matching the solution
continuously \cite{asch}. In this regime it turns out that the time of
integration, $T$, needs only to be of the order of a few $m$ in order to
obtain reliable results. On the other hand, if we try to find the solution
as a function $x(t)$ in the regime $\tau >2m$, the time of integration needs
to be of the order of a few $\tau$, which, since we are interested in large
values of $\tau$, becomes a problem. Fortunately, as we have seen, the
solution can be found as a function $y(x)$ in this case.

The results of the numerical solution are shown in Figure 2 for several
values of the mass. It is convenient not to plot the action $S(m,\tau )$
itself, but the reduced quantity
\beq\label{eq:4.25}
S_{r}(m,\tau)=\frac{S(m,\tau)-S_{0}}{S_{\infty}(\tau)}
\eeq
since this is finite in the limits $\tau \rightarrow 0$ and
$\tau \rightarrow \infty$, having the values zero and one respectively.
Here $S_0$ is the action when $\tau =0$ and $S_{\infty}(\tau )$ is the action
in the large $\tau$ limit and is given by \cite{bmn}
\beq\label{eq:4.24}
S_{\infty}=\frac{2\tau}{27}
\eeq
$S_0$ and $S_{\infty}$ are $m$ independent. For $\tau > 2m$, the hollow points
have been calculated using $(y,x)$ variables and the solid points using $(x,t)$
variables. The lines are curves of best fit through these points to aid the
eye.

This figure shows that numerically the mass has little effect on the action,
and the overdamped Langevin equation provides an excellent approximation to
the action for the underdamped system. There seems no reason to expect this
a priori, except for the fact that since in the limit of small and large $\tau$
the action is $m$ independent, there is very little freedom at intermediate
values of $\tau$ to have significant deviations from the $m=0$ result.

\section{Quasi-monochromatic noise}

In the last two sections two explicit types of noise have been considered:
white noise whose power spectrum is flat and exponentially correlated noise
which has a spectrum centered about zero. A type of noise that has a definite
color, in the sense that it has a power spectrum peaked at a non-zero
frequency, is quasi-monochromatic noise (QMN) \cite{mid}-\cite{sje}. Systems
acted upon by QMN are the subject of this section.

The noise $\xi (t)$ is defined by
\beq\label{eq:qmn}
\ddot{\xi} + 2\Gamma \dot{\xi} + \omega ^{2}_{0}\xi = \eta
\eeq
where $\eta $ is a Gaussian white noise of strength $D$. Hence for the
overdamped system $\dot{x} + V'(x) = \xi (t)$, the Lagrangian is given by
\beq\label{eq:4.36}
L(x,\dot{x},\ddot{x},\xtdots)=\frac{1}{4}\left[
(\dot{x}+V'(x))+\frac{2\Gamma}{\omega_{0}^{2}}(\ddot{x}+\dot{x}V''(x))
+\frac{1}{\omega_{0}^{2}}(\xtdots+\ddot{x}V''(x)+\dot{x}^{2}V'''(x))
\right]^{2}
\eeq
and the Hamiltonian found to be:
\beq\label{eq:4.37}
H(\bfq,\bfp)=p_{1}q_{2}+p_{2}q_{3}+\omega_{0}^{4}p_{3}^{2}
-p_{3}\left\{ \omega_{0}^{2}(q_{2}+V') +2\Gamma (q_{3}+q_{2}V'')
+q_{3}V''+q_{2}^{2}V''' \right\}
\eeq
The dynamics will be governed, in the limit of weak noise, by solutions of
Hamilton's equations given by (\ref{eq:hamq}) and (\ref{eq:hamp}). For
concreteness we will again consider the potential to be of the form
(\ref{eq:v}), and hence we will require the truncated infinite time
boundary conditions (\ref{eq:4.34}) for the uphill path. The action for
the uphill path will be given by
\beq\label{eq:4.40}
S=\int_{-\frac{T}{2}}^{\frac{T}{2}} \omega_{0}^{4}p_{3}^{2}~dt
\eeq
The downhill path again leads to zero action and will not be considered
further.

An analysis of the linearised Hamilton's equations near the end-points along
the lines described in the last section, again shows there to be oscillations
depending on the value of $\Gamma$ (in fact oscillations occur for
$\Gamma < \min (2, \omega_{0})$) and so once again we are unable to use the
$(y,x)$ parameterization of the solution. Figure 3 shows the generalized
coordinates found by solving Hamilton's equations for $\Gamma = 0.45$ and
$\omega _0 = 10$. This particular value of $\omega _0$ was chosen to allow
comparison with earlier work \cite{sje} where an approximate solution to the
classical dynamics was used to calculate the action. The value of $\Gamma $
is chosen for clarity: for smaller values it is harder to illustrate
graphically a complete transition showing the smaller scale oscillatory
features
characteristic of QMN, whereas for larger values these oscillations are
absent. The paths have three distinctive features:

\begin{enumerate}
\item{An underlying oscillatory factor of angular frequency $\omega_{0}$}.
\item{An underlying growth and decay either side of the transition time $t_{0}$
given approximately by
$\exp (-\Gamma |t-t_{0}|)$}.
\item{They pass over the top of the potential barrier many times before coming
to rest}.
\end{enumerate}

These features only occur if $\Gamma$ is less than a critical value
$\Gamma _c$ (which has a value just less than a half), otherwise the
solution is that of the system acted upon by white noise to an accuracy of
$1$\% (i.e. of order $1/\omega^2_0$). They explain why the approximate
treatment given in \cite{sje} was successful: there it was assumed that the
paths had exactly the features 1) and 2) above. The last point mentioned
above shows that one has to distinguish clearly between a mean first passage
and a well transition.

Figures 4 and 5 show the second and third figures of \cite{sje} redrawn with
the action calculated from the Hamiltonian technique shown as a dotted line.
The asterisk on the action $S$ indicates that it is the most probable escape
path (MPEP) --- for $\Gamma < \Gamma _c$ the escape path can be either
white-noise-like or oscillatory, but it is the latter that occurs in practice
since it has the least action and so is most probable. These two figures
show the remarkably good agreement between solving the full equations and
the approximation used in \cite{sje}: the value of $\Gamma _c$ is
approximately the same and a maximum value of $S^{*}/\Gamma$ occurs at
$\Gamma\sim 0.1$. From figure 4 one can also see that for $\Gamma < \Gamma _c$,
$S^{*} \approx \frac{2}{3}\Gamma$ and for $\Gamma > \Gamma _c$,
$S^{*} \approx \frac{1}{4}$. The intersection of these lines gives
$\Gamma _c = \frac{3}{8}$, which is a reasonable estimate.

Difficulties arise for small $\Gamma$ as the time required for transition
goes as $\Gamma^{-1}$ and hence a longer time span, $T$, is required. If we
attempt to rescale time by $\Gamma$, the frequency of oscillations now goes as
$\omega_{0}/\Gamma$, which means we need a finer grid of shooting points to
calculate the action to sufficient accuracy. So far we have only been able
to extend our method down to $\Gamma=0.05$.

If one writes down an equation for the optimum path (given by equation (12)
of \cite{sje}) perturbatively in powers of $1/\omega^2_0$, one find that the
uphill solution is
\beq\label{eq:4.40a}
\dot{x}=V'(x)+O\left( \frac{1}{\omega_{0}^{2}} \right)
\eeq
which has the corresponding action
\beq\label{eq:4.40b}
S=\frac{1}{4}-\frac{1}{4\omega_{0}^{2}}+O\left( \frac{1}{\omega_{0}^{4}}
\right)
\eeq
This approximate solution is independent of the value of $\Gamma$, and exists
independently of the value of $\Gamma$. For $\Gamma > \Gamma _c$ it is the
global minimum, however for $\Gamma < \Gamma_{c}$ one finds that the optimum
path from minus one to zero bifurcates. In this case this white-noise-type
path becomes a local maximum and the oscillatory-type path becomes a local
minimum. The existence of these latter paths is not obvious when solving
Hamilton's equations numerically; a very thorough search in phase space is
required to find them. This situation is common in a system such as this with
several degrees of freedom: the existence of caustics and focusing gives rise
to bifurcations in optimal paths \cite{ms1}, which makes the prediction of
the correct action difficult.

An obvious extension of this work is investigate the driving of the
underdamped Langevin equation (\ref{eq:4.13})
by QMN $\xi (t)$ given by (\ref{eq:qmn}). We might expect that driving an
equation such as this with harmonic noise such as QMN, we would find
a problem which is inherently unstable with oscillatory solutions. This
is indeed the case, and the problem has to be solved using the time as
independent variable and by use of a multiple shooting technique. Another
added problem is the introduction of two more coordinates, since the
Lagrangian now has fourth-order time derivatives. Though this does not cause
any further instabilities, it does add to the complexity of the problem and
further complicates finding the required solution. We shall not pursue this
extension any further, since it does not introduce any novel features.

\section{Conclusion}

The Hamiltonian formalism has proved effective for obtaining results for
stochastic systems governed by complicated differential equations. It has
allowed us to understand why optimum paths take on particular forms. It
also indicates that caustics and focusing appear as general features of
systems governed by colored noise (even those with only one degree of freedom)
and not just those with white noise and more than one degree of freedom. The
Hamiltonian formalism is the natural one in which to investigate and
understand these caustics systematically. We have also shown that the
technique of
of multiple shooting, though slower than relaxation and less convergent, has
allowed us to study regions which have so far been elusive and has opened up
the solution of these instanton paths in terms of the original $(x,t)$
variables. This could be useful when investigating time-dependent Lagrangians
or more complex oscillatory problems. We now feel that the structure and
general features of optimal paths are better understood, and that as a
consequence the weak-noise evaluation of escape rates and the stationary
probability distribution for many stochastic processes is now becoming
more straightforward.

\section*{Acknowledgements}

SJBE would like to thank the EPSRC for a research studentship. This work
was also supported in part under EPSRC grant GR/H40150.

\section*{Figure Captions}
\begin{enumerate}
\item{$S_{1}(m)$ plotted against $\log_{10}(m)$}
\item{$S_{r}(m,\tau)$ against $\log _{10}(\tau)$ for different values
of $m$}
\item{$q_{1}$,$q_{2}$ and $q_{3}$ against $t$ for $\Gamma=0.45$ and
$\omega_{0}=10$}
\item{Minimum QMN action $S^{*}$ against $\Gamma$.  The dotted line is from
the Hamiltonian method}
\item{Minimum QMN action $S^{*}/\Gamma$ against $\Gamma$.  The dotted line is
from the Hamiltonian method}
\end{enumerate}


%
\newpage
\begin{figure}
\begin{center}
\mbox{
\epsfysize=15cm
\epsfxsize=\textwidth
\epsfbox{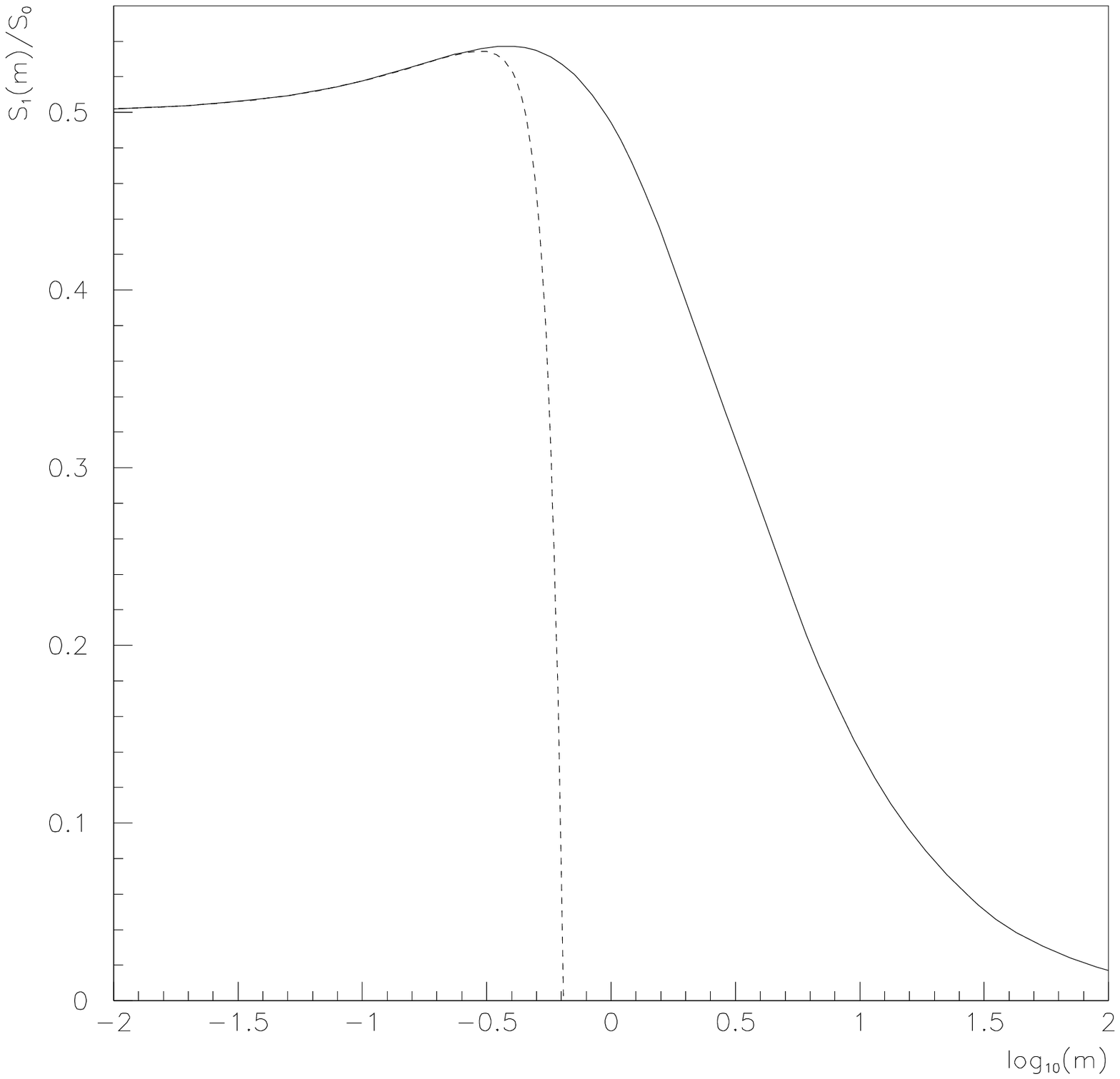}
}
\\Figure 1
\end{center}
\end{figure}
\begin{figure}[ht]
\begin{center}
\mbox{
\epsfysize=15cm
\epsfxsize=\textwidth
\epsfbox{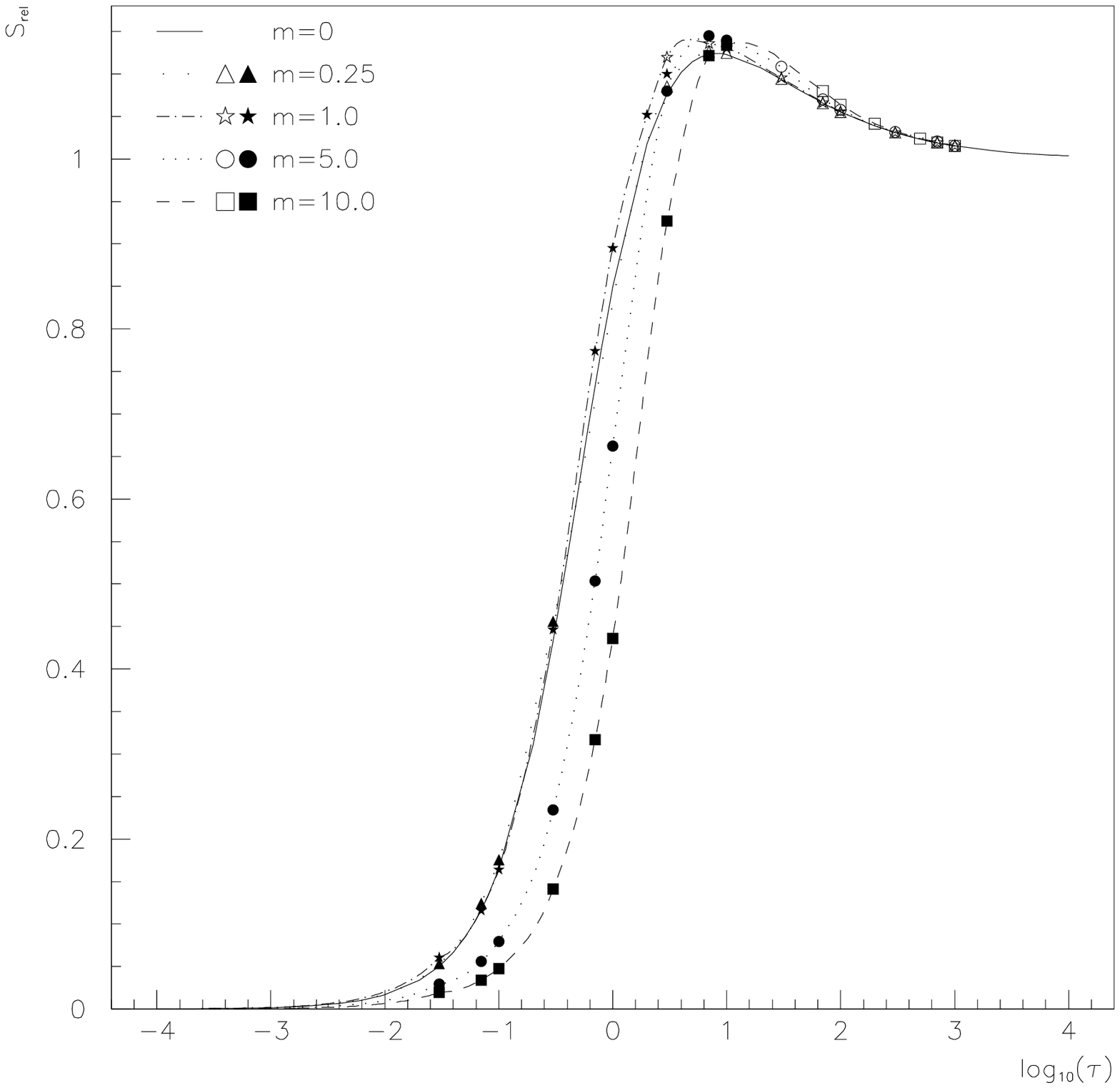}
}
\\Figure 2
\end{center}
\end{figure}
\begin{figure}
\begin{center}
\mbox{
\epsfysize=15cm
\epsfxsize=\textwidth
\epsfbox{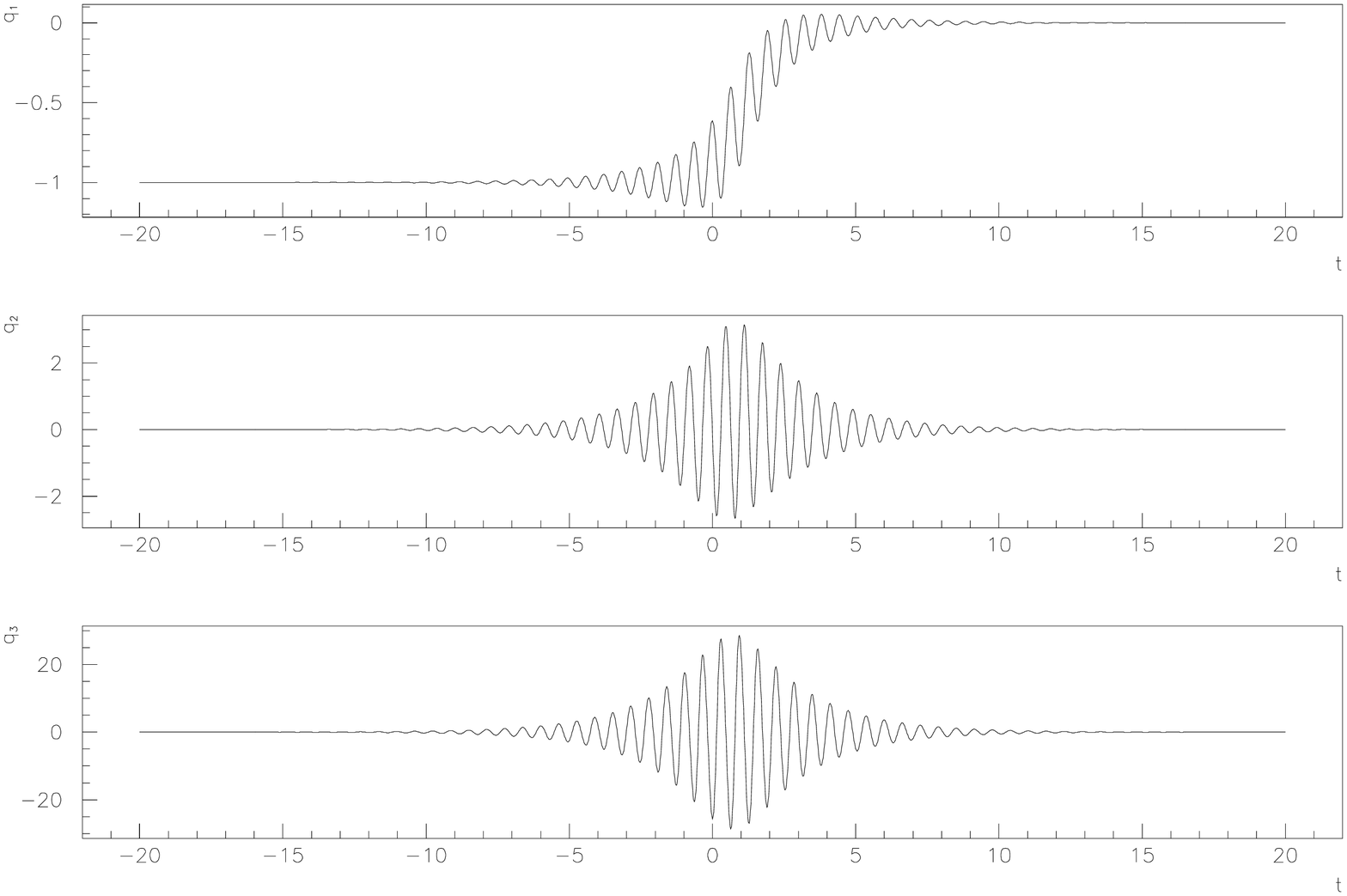}
}
Figure 3
\end{center}
\end{figure}
\begin{figure}
\begin{center}
\mbox{
\epsfxsize=\textwidth
\epsfysize=15cm
\epsfbox{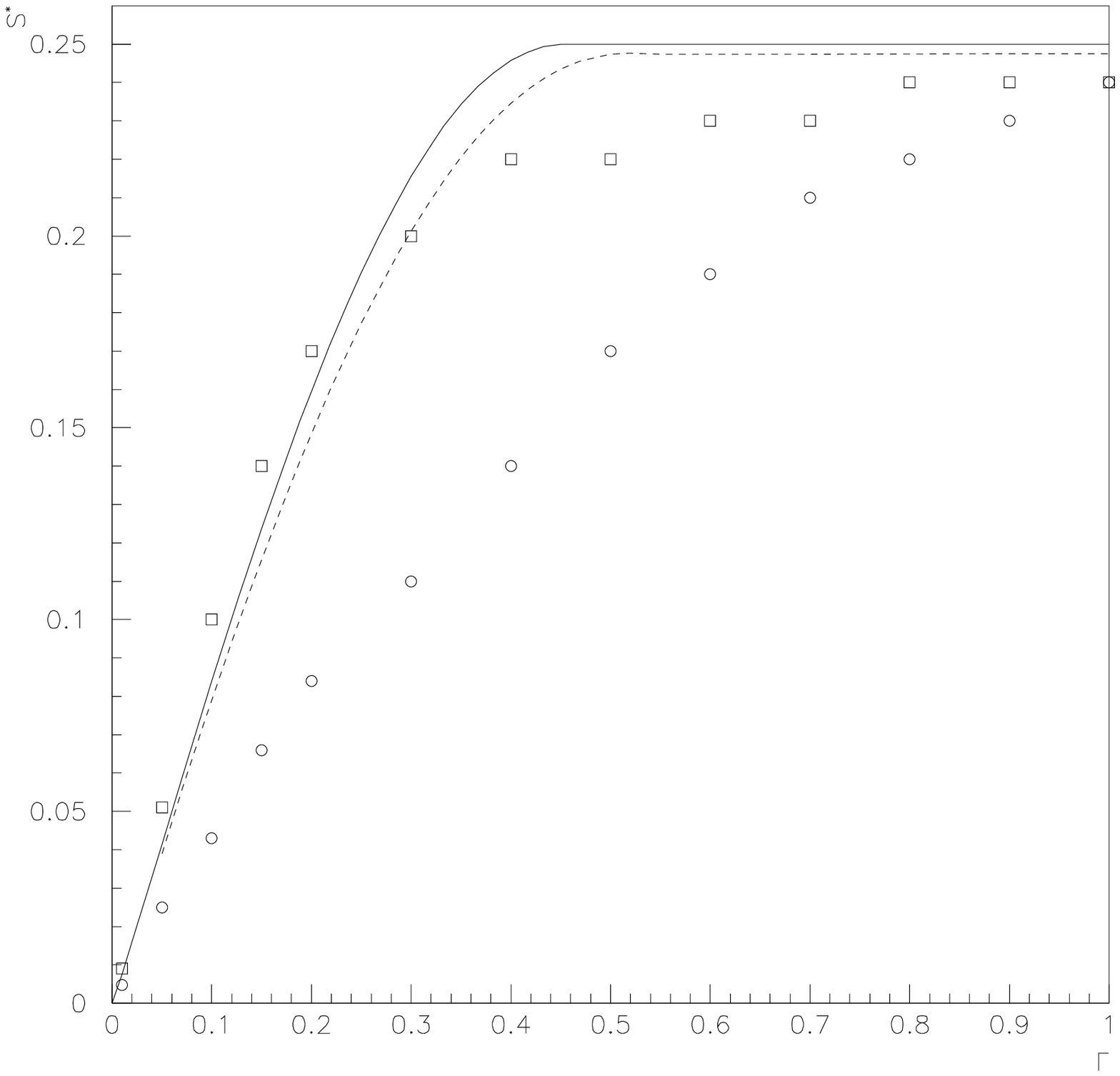}
}
Figure 4
\end{center}
\end{figure}
\begin{figure}
\begin{center}
\mbox{
\epsfxsize=\textwidth
\epsfysize=15cm
\epsfbox{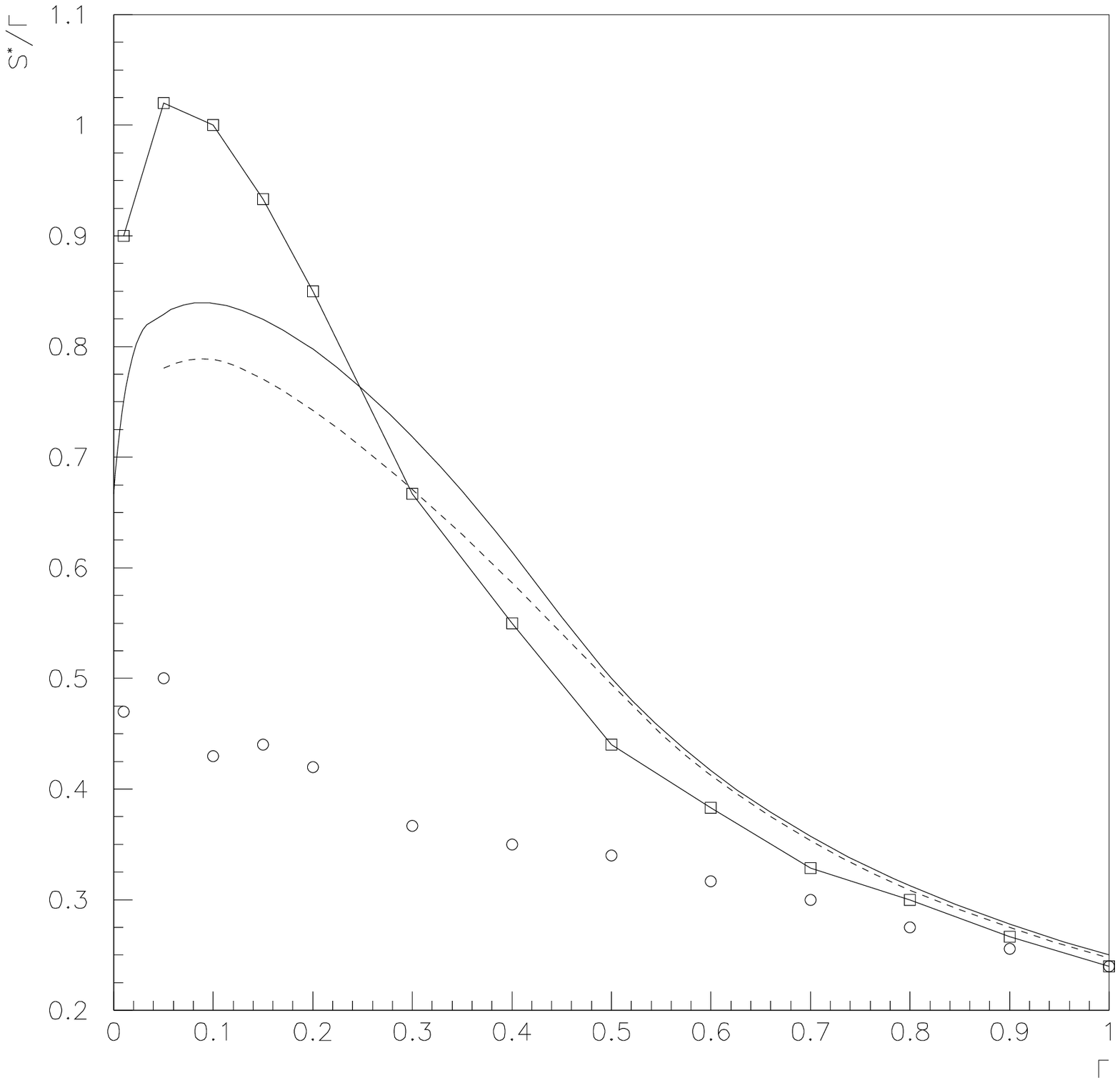}
}
Figure 5
\end{center}
\end{figure}
\end{document}